\documentclass[12pt]{article}
\usepackage[utf8]{inputenc}
\usepackage[english]{babel}
\usepackage{amssymb,graphicx}
\pdfoutput=1
\usepackage{graphics}
\usepackage{amsmath,amsfonts}
\usepackage{fullpage}

\def\e{{\rm e}}

\def\d{\partial}
\def\l{\left(}
\def\r{\right)}
\def\la{\langle }
\def\ra{\rangle }
\def\sh{\mathrm{sh}}
\def\ch{\mathrm{ch}}

\def\arcth{\mathrm{arcth}}

\newcommand{\be}{\begin{equation}}
\newcommand{\ee}{\end{equation}}
\newcommand{\bea}{\begin{eqnarray}}
\newcommand{\eea}{\end{eqnarray}}
\newcommand{\bg}{\begin{gather}}
\newcommand{\eg}{\end{gather}}
\newcommand{\bseq}{\begin{subequations}}
\newcommand{\eseq}{\end{subequations}}

\renewcommand{\ln}{\mathop{\rm ln}\nolimits}

\newcommand{\Tr}{{\rm Tr}}

\newcommand{\sla}[1]%
        {{\raise.15ex\hbox{$/$}\kern-.57em #1}}

\author{Petr Satunin\thanks{{\bf e-mail}: satunin@ms2.inr.ac.ru}
\vspace{.2cm}\\
\normalsize\it Institute for Nuclear Research of the Russian Academy of Sciences, \\ 
      \normalsize \it  60th October Anniversary Prospect, 7a, 117312  Moscow, Russia} 

\title{Width of photon decay in magnetic field:\\ elementary semiclassical
  derivation and sensitivity to Lorentz violation}
\date{}
\begin{document}
\maketitle

\begin{abstract}

We present an elementary derivation of the width of photon decay 
in a weak magnetic field using the semiclassical method of
worldline instantons. The calculation is generalized to a model of 
quantum electrodynamics with broken Lorentz symmetry. Implications for
the search of deviations from Lorentz invariance in the cosmic
ray experiments are discussed.

\end{abstract}

\section{Introduction}

$\mbox{\quad\ }$
Semiclassical methods are widely used in modern quantum field
theory. They provide a powerful tool to investigate non-perturbative
phenomena. The well-known example is the false vacuum decay
\cite{Kobzarev:1974cp,Coleman:1977py}. Equations of motion have a
solution, called bounce, that interpolate between false and true
vacua. The probability of decay is proportional to the exponent (with
the minus
sign) of the Euclidean action evaluated on the bounce. 

The similar calculation arises in another class of processes, particle
production in external backgrounds. The simplest example is the Schwinger
effect --- spontaneous creation of electron-positron pairs in external
electric field. The probability of pair production is expressed
through the thermal partition function of a certain quantum mechanical
problem \cite{Affleck:1981bma}. This partition function is evaluated
in the saddle point approximation;
the solutions  of the saddle point equations, called 'worldline instantons',
are interpreted as trajectories of a particle in an auxiliary periodic
time. 
Semiclassical treatment of Schwinger-like processes has been
generalized to the cases of time-dependent and inhomogeneous
electromagnetic field \cite{Dunne:2005sx} and to the photon-stimulated
Schwinger pair creation in application to the laser physics
\cite{Monin:2010qj}. In this article we study the similar process ---
the photon decay into an electron-positron pair in magnetic field. 
We will be interested in the weak-field limit where the semiclassical
method is applicable. For a review of methods used in the opposite
case of the strong magnetic field see~\cite{Kuznetsov:2004tb}. 

The probability of the photon decay in magnetic field was calculated
long time ago independently by \cite{Robl,Klepikov}. The
corresponding matrix
element has been computed in the semiclassical
approximation in terms of the overlap of electron wavefunctions
(eigenfunctions of the Dirac equation in the uniform magnetic field)
in the 
coordinate representation. In the weak-field limit the
photon decay is exponentially
suppressed,
\be\label{intro} 
\Gamma \propto \exp\l -\frac{8m^3}{3\omega eH\sin\varphi}\r.
\ee
Here $\omega$ is the energy of the photon, $H$ is the 
value of the uniform magnetic field, $\varphi$ denotes the angle
between the photon momentum and the 
magnetic field, $e$ and $m$ are the electron charge and 
mass; it is assumed that the photon
energy is much higher than the electron mass, $\omega\sin\varphi\gg m$
(but still $\omega e H \sin\varphi\ll m^3$). 
The calculation of \cite{Robl,Klepikov} is technically
quite involved. The approach based on the 'worldline instantons'
adopted in the present paper is significantly simpler and has a clear
geometrical interpretation. To the best of our knowledge, it has not
been applied to the process of pair production in magnetic field so far.

Due to its simplicity, our method can be easily generalized to models
beyond the standard QED. This is illustrated in the second part of the
paper where we study the process of gamma decay in magnetic field in a
model of electrodynamics without
Lorentz invariance (LI). It is possible that deviation from LI can
appear at very high energies, unaccessible to present-day
accelerators. This scenario is suggested by several approaches to the
theory of quantum gravity
\cite{Ellis:2003if,Mavromatos:2010pk,Girelli:2012ju,Horava:2009uw,Blas:2009qj}
and the energy scale where deviations from LI become significant is
naturally assumed to lie at the Plank mass $M_p$ or a few orders
below. 

Remarkably, this type of Lorentz violation (LV) can be constrained by
cosmic ray observations, see \cite{Liberati:2012th} for recent
review. 
Indeed, energies attained by particles of
ultra-high energy cosmic rays (UHECR) greatly exceed those achieved in
laboratory. Reaching the Earth, cosmic ray primaries interact in the
atmosphere and create showers of descendant particles, that can be
detected experimentally. The characteristics of the shower
depend on the type and energy of the primary. 

The process of photon decay in magnetic field plays an important role
in these consideration.  
When an UHECR photon (with energy above $10^{19.5}\,\mbox{eV}$)
reaches the magnetic field of the Earth, it decays into an
electron and positron. The latter, in turn, produce photons by the
synchrotron radiation, giving rise to an electromagnetic cascade. 
In this way the magnetosphere shower called preshower
\cite{Homola:2003ru,Risse:2007sd} is created at the altitude of
several hundred kilometers above the Earth surface. 
Preshower accelerates the subsequent shower development in the
atmosphere and provides a unique signature for photon-induced
showers: pair production probability depends on the perpendicular
component of the magnetic field and hence on the arrival direction of
the photon. We will find that possible LV significantly affects the
probability of the preshower formation. Thus, experimental detection
of a preshower would provide a sensitive probe of LV.

The paper is organized as follows. In Sec.~\ref{sec:2} we derive
Eq.~(\ref{intro}) by the semiclassical method of worldline
instantons. In Sec.~\ref{sec:3} we calculate the width of photon decay in
magnetic field in QED with LV. Sec.~\ref{sec:4} is devoted to discussion. 


\section{Photon decay in a weak magnetic field in standard QED}
\label{sec:2}

Consider a photon with
four-momentum $k_\mu=(\omega,{\bf k})$ propagating in the uniform
magnetic field ${\bf H}$ at an angle $\varphi$ to its direction. We
choose the coordinate system where the magnetic field points along
the $x$-axis, ${\bf H}=(H,0,0)$, and the spatial photon momentum lies
in the $(x,y)$-plane,
${\bf k}=(\omega\cos\varphi,\omega\sin\varphi,0)$. The photon decay
into an $e^+e^-$ pair is
kinematically allowed if $\omega\sin\varphi > 2m$. We will assume the
photon energy to be well above this threshold, 
$\omega\sin\varphi \gg m$. 

To find the rate of the photon decay we adopt the method similar to
that used in \cite{Affleck:1981bma,Dunne:2005sx,Monin:2010qj} for the
semiclassical analysis of the Schwinger process. It was shown in these
works that in the leading approximation the answer is insensitive to
the spin of the electron. 
As in the present 
paper we are interested only in the leading order result, we choose to
work for simplicity with the scalar QED, described by the
Lagrangian\footnote{We take the signature $(+,-,-,-)$ for the
  Minkowski metric.} 
\be\label{scalarL}
\mathcal{L} = -\frac{1}{4}F_{\mu\nu}F^{\mu\nu} + D_\mu\phi^* D^\mu\phi
-m^2\phi^*\phi, 
\ee
where the covariant derivative $D_\mu$ is defined in the usual way,
$D_\mu\phi=\l\d_\mu-ieA_\mu\r\phi$.  

It follows from the optical theorem that the rate of photon decay is
proportional to the imaginary part of the polarization operator 
\be\label{OpticalTH}
\Gamma = \frac{1}{2\omega} \epsilon_\mu(k)\epsilon_\nu(k)
\mbox{Im}\Pi_{\mu\nu}(k),
\ee
where $\epsilon_\mu(k)$ is the photon polarization vector which we
choose to be real. As usual, 
$\Pi_{\mu\nu}$ is given by the Fourier transform of the 
correlator of two electromagnetic currents
$j_\mu=-ie(\phi^*D_\mu\phi-\phi D_\mu\phi^*)$,
\be
\label{polar0}
\Pi_{\mu\nu}(k) =\int d^4 y\, \e^{iky}\la j_\mu(y/2)j_\nu(-y/2)\ra\;.
\ee
In the leading order one can neglect the contribution of virtual
photons into this correlator which is then
expressed in terms of the partition function of the charged scalar
in external electromagnetic field,  
\be\label{polar}
\la j_\mu(y/2)j_\nu(-y/2)\ra = \frac{1}{Z[A_\mu]}\frac{\delta}{i\delta A_\mu(y/2)}\frac{\delta}{i\delta A_\nu(-y/2)} Z[A_\mu]\;,
\ee
where
\be\label{Z}
Z[A_\mu]=\int D\phi^*D\phi\, \e^{-\int d^4x\l \left|D_\mu\phi\right|^2+m^2\left|\phi\right|^2\r}= \det\l -D_\mu^2 + m^2\r = \exp \Tr\ln \l -D_\mu^2 + m^2\r\;.
\ee
Note that in defining the partition function we have 
performed the Wick rotation to the Euclidean signature. 

At the next step we use the formula:
$$
-\ln M = \int_0^\infty\frac{dT}{T}\l \e^{-MT} - \e^{-T}\r\;.
$$ 
This leads to the expression for the partition function in terms of
the integral over the ``proper time'' $T$. 
\be\label{zz0}
Z[A_\mu]=Z_0\exp\left[ - \int_0^\infty\frac{dT}{T}\e^{-m^2T} \Tr
  \l \e^{TD_\mu^2}\r \right].
\ee
The operator $\l -D_\mu^2\r$ can be interpreted as the
quantum-mechanical Hamiltonian of a point particle in
four-dimensional space. Thus, $ \Tr\l \e^{-T\l-D_\mu^2\r}\r$ can be
considered as its thermal partition function with the proper time $T$
playing the role of the inverse temperature. 
It is more convenient to work in the Lagrangian formalism, so we make
the Legendre transformation and consider the functional integral
representation: 
$$
\Tr \l \e^{TD_\mu^2}\r = \int\limits_{p.b.c.}Dx_\mu
\e^{-\int_0^T d\tau\l \frac{\dot{x}^2_\mu}{4} + ie\dot{x}_\mu
  A_\mu\r}. 
$$
Here we have introduced an auxiliary time $\tau$ and the notation
$p.b.c.$ means periodical boundary conditions
$x_\mu(0)=x_\mu(T)$. The partition function becomes 
\be
Z[A_\mu]=Z_0\exp\left[ - \int_0^\infty\frac{dT}{T}\e^{-m^2T}\int_{p.b.c}Dx_\mu\,\e^{-\int_0^T d\tau\l \frac{\dot{x}^2_\mu}{4} + ie\dot{x}_\mu A_\mu\r}\right].
\ee

We now return to the polarization operator (\ref{polar0}).
Each variational derivative of the partition function with respect to
$A_\mu$ produces an insertion of the combination $\oint d\tau
\dot{x}_\mu(\tau)\delta(x(\tau)-y)$ in the functional
integral. Also we rescale the auxiliary time $\tau$ to make it vary
from $0$ to $1$. This gives, 
\begin{align}
\la j_\mu(y/2)j_\nu(-y/2)\ra \propto &\int_0^\infty\frac{dT}{T}\int_{p.b.c.} Dx_\mu \oint d\tau_1\oint d\tau_2 \dot{x}_\mu(\tau_1)\dot{x}_\nu(\tau_2)\nonumber
\\&\times\delta(x(\tau_1)-y/2)\delta(x(\tau_2)+y/2)\;
\e^{-m^2T-\int_0^1 d\tau\l \frac{\dot{x}^2_\mu}{4T} + ie\dot{x}_\mu A_\mu\r}.\nonumber
\end{align}
The width of the photon decay is given by the imaginary part of the
polarization operator in the momentum representation
$\mathrm{Im}\,\Pi_{\mu\nu}(k)=\mathrm{Im}\int
d^4y\,e^{iky}\Pi_{\mu\nu}(y)$. Taking the Fourier transform for the imaginary part of the polarization
operator\footnote{Strictly speaking, the integral here should be
  performed over the configurations satisfying
  $x(\tau_1)=-x(\tau_2)$. We will ignore this restriction because it does not affect the final result.}, one obtains
\be\label{PiMuNu}
\mathrm{Im}\,\Pi_{\mu\nu}(k)\propto\mathrm{Im}\,\int_0^\infty\frac{dT}{T}\int_{p.b.c.}
Dx_\mu \oint d\tau_1\oint d\tau_2 \,
\dot{x}_\mu(\tau_1)\dot{x}_\nu(\tau_2)\e^{-S_m[x_\mu;\tau_1,\tau_2]},
\ee
where
\be\label{actionQM}
S_m[x_\mu;\tau_1,\tau_2]=m^2T+\int_0^1 d\tau\l \frac{\dot{x}_\mu^2}{4T}+ieA_\mu\dot{x}_\mu\r-ik_\mu\l x_\mu(\tau_1)-x_\mu(\tau_2)\r.
\ee
This expression has the form of the Euclidean action of a relativistic
particle in the external electromagnetic field with two sources of
opposite signs located at the proper times $\tau_1$ and $\tau_2$. The
strength of these sources is determined by the photon
momentum. Integrating out the parameter $T$ one can obtain the standard
form of the relativistic particle action. 

\begin{figure}
\centering
\includegraphics{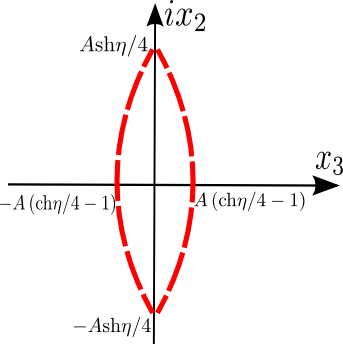}
\caption{
\label{fig:trajectory}
'Worldline instanton' --- the classical trajectory describing
  pair production in the magnetic field for the case when the photon
  momentum is perpendicular to the field direction, $\varphi=\pi/2$. 
The projection of the trajectory on the plane ($ix_2,\,x_3$) is shown.} 
\end{figure}

We will evaluate the r.h.s. of (\ref{PiMuNu}) in the saddle-point
approximation. To this end, firstly we find the saddle 
equations for $T$ and $x_\mu(\tau)$. Their solution gives the
saddle-point classical trajectory $x_\mu^{cl}(\tau)$. At the second
step the trajectory is substituted into the action (\ref{actionQM}). 
Let us fix the gauge $A_\mu=-\frac{1}{2}F_{\mu\nu}x_\nu$. Varying over
$x_\mu$, we obtain (separately for the time and space components of
$x_\mu$),  
\begin{align}
&\frac{\ddot{x}_0}{2T} = \omega\l \delta\l\tau-\tau_1\r - \delta\l\tau-\tau_2\r \r,\label{e1}\\
&\frac{\ddot{x}_i}{2T} - ieF_{ij}\dot{x}_j = - i\omega\left[\delta_{i1}\cos\varphi + \delta_{i2}\sin\varphi\right] \l \delta\l\tau-\tau_1\r - \delta\l\tau-\tau_2\r \r. \label{e2}
\end{align}
The variation over $T$ yields,
\be\label{saddleT}
m^2 - \frac{\int_0^1 d\tau \dot{x}_\mu^2}{4T^2}=0.
\ee 
We are looking for a solution of Eqs.~(\ref{e1})--(\ref{saddleT}) that
describes a closed trajectory in 4-dimensional spacetime. Solutions of
this type are called ``worldline instantons''. Note that in general they
can be complex (cf. \cite{Monin:2010qj}). The solution exists if
$|\tau_1-\tau_2|=\frac{1}{2}$. Without loss of generality we set $\tau_1=0,\
\tau_2=\frac{1}{2}$. Our solution is composed of two hyperbolic arcs
(see Fig.~\ref{fig:trajectory}) defined on the segments $\tau \in
(0,\frac{1}{2})$ and $\tau \in
(\frac{1}{2},1)$ respectively,
\begin{align}
\text{for~~} 0<\tau<\frac{1}{2}:&\qquad
x_0^{cl}=\omega T\l \tau - \frac{1}{2}\r,&
x_1^{cl}=-iA\eta\cos\varphi\,\ch\frac{\eta}{4}\cdot\l\tau-\frac{1}{4}\r, \label{sol1}\\
x_2^{cl}=-iA&\sin\varphi\,\sh\l\eta\l\tau-\frac{1}{4}\r\r,&
x_3^{cl}=-A\sin\varphi\,\left[\ch\l\eta\l\tau - \frac{1}{4}\r\r -\ch\frac{\eta}{4} \right];\notag\\
\text{for~~} \frac{1}{2}<\tau<1:&~~~~~
x_0^{cl}=-\omega T\l \tau - \frac{1}{2}\r,&
x_1^{cl}=iA\eta\cos\varphi\,\ch\frac{\eta}{4}\cdot\l\tau-\frac{3}{4}\r, \label{sol2}\\
x_2^{cl}=iA&\sin\varphi\,\sh\l\eta\l\tau-\frac{3}{4}\r\r,& 
x_3^{cl}=A\sin\varphi\,\left[\ch\l\eta\l\tau - \frac{3}{4}\r\r -\ch\frac{\eta}{4} \right].\notag
\end{align}
Here parameters $A,\,\eta$ are determined from Eqs.~(\ref{e1}), (\ref{e2}):
\be\label{Aeta}
A=\frac{\omega}{2eH\ch\frac{\eta}{4}}, \qquad \eta=2TeH.
\ee
Substituting the solution (\ref{sol1})--(\ref{Aeta}) into
(\ref{saddleT}) one obtains\footnote{This formula is valid in the regime $\omega\sin\phi \gg 2m$. The exact expression reads $T=\frac{2}{eH}\arcth\frac{2m}{\omega\sin\varphi}$.}:   
\be\label{TLI}
T = \frac{4m}{\omega eH\sin\varphi}.
\ee

The next step is to evaluate the action (\ref{actionQM}) on the
solution. After a straightforward calculations we obtain: 
\be\label{LIanswer}
S[x_\mu^{cl}]=\frac{8}{3}\frac{m^3}{\omega eH\sin\varphi}.
\ee
The semiclassical method is valid as long as the classical action is
large\footnote{As an example, let us consider the geomagnetic field 
$H \sim 0.5\,G$ and take $\sin\varphi \sim 1$. Then the method is
applicable for photons with the energy $\omega \lesssim
10^{19.5}\,\mathrm{eV}$.},  
$S[x_\mu^{cl}] \gg 1$.  
Combining everything together, we obtain the 
photon decay width\footnote{In general, one should sum over
  contributions of all possible classical solutions, we consider only
  the dominant one.}:
$$
\Gamma = \frac{1}{2\omega} 
\epsilon_\mu(k)\epsilon_\nu(k)\mathrm{Im}\l\dot{x}_\mu^{cl}(0)\dot{x}_\nu^{cl}(1/2)\mathcal{N}e^{-S[x_\mu^{cl}]}\r,
$$
where $\dot{x}_\mu^{cl}$ is the derivative of the classical
solution\footnote{Note that the discontinuity of $\dot x^{cl}_\mu$ at
  $\tau=0,1/2$, implied by Eqs.~(\ref{e1}), (\ref{e2}), is
  proportional to $k_\mu$ and thus vanishes when contracted with the
  polarization vector.}
and $\mathcal{N}$ is a pre-exponential factor coming from the
integration over fluctuations near the classical solution
(\ref{sol1})--(\ref{Aeta}). Using the reasoning similar to
\cite{Affleck:1981bma} one can show that the prefactor $\mathcal{N}$
has a single negative mode, corresponding to the variation of the size
of the worldline instanton. Therefore, according to the standard
arguments \cite{Coleman:1977py} the prefactor is imaginary and the
decay width is nonzero. In our approximation we are interested only in
the leading exponential behavior. Thus we neglect the prefactor and
arrive at the formula (\ref{intro}).

\section{Generalization to QED with Lorentz violation}
\label{sec:3}

The simplicity of the method presented in the previous section allows
us to easily generalize it to non-standard theories. In this section
we analyze the sensitivity of the photon decay width in the magnetic
field to possible deviations from Lorentz invariance. To make the
calculation concrete we need to specify the model. We consider the
analog of the model introduced in \cite{Rubtsov:2012kb}, where we
replace the spinor electron field by a charged scalar.  
The Lagrangian reads,
\begin{gather}\label{LVLagr}
\begin{align}
\mathcal{L} = &-\frac{1}{4}F_{\mu\nu} F^{\mu\nu} + D_\mu\phi^*D^\mu\phi - m^2\phi^*\phi \,+ \\
&  + \frac{1}{4}F_{kj}\l -2\varkappa+\frac{\xi\d_i^2}{M^2} \r F^{kj} - \frac{2g}{M^2}D_i^2 \phi^*D_i^2\phi.\notag
\end{align}
\end{gather}
Here the first line represents the standard LI scalar QED, while the second line
contains extra LV operators of dimension 4 and 6; $\varkappa,\,\xi$
and $g$ are dimensionless 
coefficients, $M$ is a parameter of order the Planck mass. The form of
the Lagrangian is fixed by requiring the theory to be 
rotationally invariant in the
preferred frame, gauge invariant, and CPT- and P-even. The full list
of restrictions on the theory and their motivation are discussed in
\cite{Rubtsov:2012kb}.
  
From (\ref{LVLagr}) one obtains the dispersion relations for photons
and (scalar) electrons and positrons, 
\begin{align}
&\gamma: \qquad \omega^2 = k^2(1-2\varkappa) + \frac{\xi k^4}{M^2},\label{DRphoton}\\
&e^{\pm}: \qquad E^2=m^2+p^2+\frac{2gp^4}{M^2}.\label{DRelectron}
\end{align}
These differ from the standard case, and as a consequence the kinematics of
various reactions is modified. In particular, the reactions relevant for
propagation and detection of the cosmic rays are affected which makes
the cosmic ray experiments sensitive to LV.

As discussed in \cite{Rubtsov:2012kb}, in general there are other
consequences of LV that may also be important. Thus, the 
sums over polarizations entering the calculation of reaction rates are
also modified; new interaction vertices appear from the last term in
the Lagrangian (\ref{LVLagr}). However, these modifications affect
only the pre-exponential factor in the polarization operator. As we
are interested only in the leading exponential behavior, we neglect
this type of correction in what follows.

The optical theorem is based only on unitarity and does not rely on
LI. Thus, to calculate the photon decay rate in magnetic field we can
still use the formula (2). As before, we neglect the contributions of
virtual photons, so that the polarization operator 
$\Pi_{\mu\nu}$ is again related
to the second variational derivative of the partition function $Z[A_\mu]$. 
On the other hand, the latter is modified due to LV. In the proper
time representation it reads,
\be\label{ZLV}
Z[A_\mu]=Z_0\exp\left[ - \int_0^\infty\frac{dT}{T}\e^{-m^2T} \Tr \e^{T\l D_\mu^2 - \frac{2g}{M^2}(D_i^2)^2\r}\right]\;,
\ee
where we have rotated to the Euclidean time.
Here the additional term $-\frac{2g}{M^2}(D_i^2)^2$ in the inner exponent
comes from the four-derivative electron kinetic term in the second
line of (\ref{LVLagr}). The next steps are the same as in
Sec.~\ref{sec:2}. One interprets the trace in (\ref{ZLV}) as a quantum
mechanical statistical sum and writes the functional integral
representation for it in terms of the point-particle action. This
leads to the expression (\ref{PiMuNu}) for the imaginary part of the
polarization operator, where now
\be\label{LVPiMuNu}
S_m=m^2T + \int_0^1 d\tau\l
\frac{\dot{x}_\mu^2}{4T}+ieA_\mu\dot{x}_\mu+
\frac{g(\dot{x}_i^2)^2}{8T^3M^2}\r - ik_\mu\l
x_\mu(\tau_1)-x_\mu(\tau_2)\r\;. 
\ee
In deriving this formula we have assumed,
\be\label{co}
\frac{g\dot{x}_i^2}{M^2T^2} \ll 1.
\ee
We will see later, that this condition is equivalent to the
requirement that the LV corrections to the electron dispersion
relation 
(the last term in (\ref{DRelectron})) is small compared to $p^2$. Note
that this still allows the LV correction to be of order or larger than
the (squared) electron mass $m^2$. Eq.~(\ref{LVPiMuNu}) differs
in two respects from the LI case (\ref{actionQM}).
First, the term $\frac{g(\dot{x}_i^2)^2}{8T^3M^2}$ appears in the
point-particle Lagrangian. 
Second, the momentum $k_\mu$ of the initial photon must satisfy
dispersion relation (\ref{DRphoton}). 

We now evaluate the integrals over $T$ and over $x_\mu$ in (\ref{PiMuNu})
by the saddle point method. Varying the action (\ref{LVPiMuNu}) over T
we obtain, 
\be\label{eLVT}
m^2 = \frac{\dot{x}_\mu^2}{4T^2} + \frac{3g(\dot{x}_i^2)^2}{8M^2T^4}.
\ee
Variation over $x_\mu$ gives the equations of motion. The
time-component of the equations does not change compared to the LI
case, see Eq.~(\ref{e1}). On the other hand, the spatial equations get
modified,\footnote{Note that on the l.h.s. of (\ref{LVe2}) we have
  omitted the term
  $\frac{g\dot{x}_i}{2T^3M^2}\frac{d\dot{x}_j^2}{d\tau}$, as 
$\dot{x}_j^2$ is the integral of
  motion.} 
\be\label{LVe2}
\frac{\ddot{x}_i}{2T}\left[ 1+\frac{g\dot{x}_j^2}{M^2T^2} \right] - ieF_{ij}\dot{x}_j = - ik\left[\delta_{i1}\cos\varphi + \delta_{i2}\sin\varphi\right] \l \delta(\tau-\tau_1) - \delta(\tau-\tau_2)\r.
\ee
The solution of Eqs.~(\ref{e1}), (\ref{LVe2}) has the form
(\ref{sol1}), (\ref{sol2}), but with different parameters $A$ and $\eta$,
\be\label{LVcond}
A=\frac{k}{2eH\ch\frac{\eta}{4}}, \qquad \eta=2TeH\left[ 1+\frac{g\omega^2}{M^2}\right].
\ee
Substituting this into eq. (\ref{eLVT}) and solving it with respect to $T$ we obtain,
\be\label{LVT}
T=\frac{2\sqrt{\l\frac{2m}{\omega}\r^2 + \frac{g\omega^2}{2M^2}-\frac{\xi\omega^2}{M^2}+2\varkappa}}{eH\sin\varphi}.
\ee
Note that none of the terms under the square root can be
neglected. Coming back to the condition 
(\ref{co}), it is now straightforward to check that on the solution it
reduces to 
$$
\frac{g\omega^2}{M^2} \ll 1.
$$
As advocated before,
this is nothing but the requirement that the ratio between the third
and the second term on the r.h.s. of (\ref{DRelectron}) is small for
the real electrons (positrons) produced in the photon decay. Clearly,
this is satisfied for all astrophysically relevant photon energies
$\omega$ if $g$ is not much bigger than 1.

Next we substitute the solution
(\ref{sol1}), (\ref{sol2}), (\ref{LVcond}), (\ref{LVT}) into the
action (\ref{actionQM}) and obtain, 
\be\label{LVe}
S[x_\mu^{cl}]=\frac{\omega^2}{3eH\sin\varphi}\l \l \frac{2m}{\omega}\r^2 + \frac{g\omega^2}{2M^2}-\frac{\xi\omega^2}{M^2}+2\varkappa\r^{3/2}.
\ee
Following \cite{Rubtsov:2012kb} we introduce the combination
$\omega_{LV}$ that characterizes the LV contribution into the
kinematics of the reaction,
\be
\label{omegaLV}
\omega_{LV}=-\varkappa\omega+\frac{\xi \omega^3}{2M^2} - \frac{g\omega^3}{4M^2}.
\ee
Using this notation the width of the photon decay is cast into the form,
$$
\Gamma \propto \exp\left[-\frac{8m^3}{3\omega eH\sin\varphi}\l 1 - 
\frac{\omega\cdot\omega_{LV}}{2m^2}\r^{3/2}\right].
$$
This is the main result of this section. Let us analyze it. Even small
negative $\omega_{LV}\lesssim-\frac{2m^2}{\omega}$ exponentially suppresses
the width of the photon decay in magnetic field. On the other hand, even small
positive $\omega_{LV}\sim\frac{2m^2}{\omega}$ decreases the absolute
value of the exponent and the decay becomes unsuppressed (of course,
the semiclassical approximation breaks down in this case, cf. the
discussion at the end of Sec.~\ref{sec:2}).

Our result admits the following interpretation. Consider for
simplicity the case when LV is present only in the electron sector, 
$\varkappa=\xi=0$. Introduce the effective momentum-dependent electron
mass by the formula 
\be\label{meff}
m^2_{eff}(p)\equiv E^2-p^2 = m^2 + \frac{2gp^4}{M^2}.
\ee
In terms of this notation the formula for the pair-production width takes
the standard form (\ref{intro}) with $m$ replaced by
$m_{eff}(\omega/2)$ --- the effective mass of the produced electron
(positron). The larger the effective mass, the more suppressed is the
photon decay, and vice versa. One concludes that in the leading
approximation the effect of LV on this process is completely
encompassed by the kinematics.

Before finishing this section let us discuss how our calculation is
modified in the case of a more general pattern of LV. Generalization
to an arbitrary LV in the photon sector is straightforward: the only
change amounts to a different relation between 
$\omega$ and $k$ in Eqs.~(\ref{e1}) and (\ref{LVe2}). The
form of the equations remains the same, and by literally repeating the
above calculation one can analytically find the suppression exponent 
in a model with arbitrary dispersion relation
$\omega(k)$.  

More technical difficulties arise if we allow for an arbitrary
dispersion relation for electrons (and the same for positrons). The saddle point equation
(\ref{eLVT}) changes and, in general, cannot be solved analytically:
all terms in this equation are comparable, so we cannot perform an
expansion in small parameter. Thus, for a general electron dispersion relation
one can find the 
width of photon decay only numerically. However, it appears on the
physical grounds that, at least qualitatively, the leading effect of
LV will be again encompassed by the kinematics. Thus, a faithful
estimate of the suppression exponent can be obtained by substituting
the effective mass $m_{eff}$ defined by the first equality in
(\ref{meff}) in the standard formula (\ref{intro}).

\section{Discussion}
\label{sec:4}

We have studied the process of photon
decay into an electron-positron pair in external magnetic field by the
semiclassical method of 
worldline instantons. 
A technically simple derivation of the leading exponential behavior of
the width of this process has been presented. 

We have shown that the method can be easily
generalized to the extension of QED including possible deviations from
the Lorentz invariance and illustrated this by an explicit 
analytical calculation in the model of LV scalar QED with dispersion relations
quartic in momentum. 
We also discussed how this calculation can be
in principle generalized to the case of arbitrary dispersion relations. 
It was found that the width of photon decay in weak magnetic field is
exponentially sensitive to the LV contributions.

Let us discuss the implications of our results for the tests of LV
involving cosmic ray observations. To date, no UHE photons with
energy $10^{19.5}\,\mbox{eV}$ or more have been detected. However,
there are reasons to expect non-zero flux of photons at
such high energies. The break in the spectrum of cosmic ray hadrons at
energy $\sim 10^{19.6}\,\mbox{eV}$ has been detected independently in
three experiments
\cite{Abbasi:2007sv,Abraham:2008ru,AbuZayyad:2012ru}. If
the dominant fraction of the cosmic ray primaries are protons
this break is naturally identified with the GZK cut-off
\cite{Greisen:1966jv,Zatsepin:1966jv}: suppression of the proton flux
due to the interactions with the 
cosmic microwave background. In this interaction multiple charged and
neutral pions are produced.
Neutral pions, in turn, decay to photons called
'cosmogenic', or 'GZK' photons
\cite{Gelmini:2005wu,Hooper:2010ze}. The
cosmogenic photons may be detected within the current decade by the
Pierre Auger 
Observatory \cite{Alvarez}.   

Imagine now the situation that a non-zero flux of photons with
energies $\omega\sim 10^{20}\,\mbox{eV}$ has been observed.
Imagine moreover that the threshold for the preshower formation in the
Earth magnetosphere has been measured to coincide with the predictions
of the standard Lorentz invariant QED. This would imply that the
quantity $\omega_{LV}$ defined in (\ref{omegaLV}) satisfies
$|\omega_{LV}|<2m^2/\omega$, or numerically  
$\left|\omega_{LV}\right|\,<\,10^{-8}\,\mbox{eV}$. 
Barring accidental cancellations between various terms entering
(\ref{omegaLV}) and taking $M=10^{19}$ GeV this would translate into
the stringent bounds $|\varkappa|<10^{-28}$;~ 
 $|\xi|, |g|<10^{-11}$. The ability to constrain $\xi$, $g$ at the
 level well below 1 means that the preshower formation by UHECR
 photons is sensitive even to trans-Planckian breaking of LI.

\paragraph*{Acknowledgements}

The author thanks Sergei Sibiryakov, Grigory Rubtsov and Alexander Monin for helpful
discussions. This work was supported in part by the Grant of the
President of Russian Federation NS-5590.2012.2, the Grant of the
Ministry of Education and Science No. 8412 and by the RFBR grants
11-02-01528, 12-02-01203, 12-02-91323.

\end{document}